\documentclass[twocolumn,showpacs,aip,cha,preprintnumbers,amsmath,amssymb,
superscriptaddress,reprint]{revtex4-1}
\usepackage{graphicx}
\usepackage{enumitem}
\usepackage{booktabs}

\makeatletter
\def\namedlabel#1#2{\begingroup #2%
	\def\@currentlabel{#2}%
	\phantomsection\label{#1}
\endgroup}
\makeatother
\newcommand{\be}{\begin{equation}}              
\newcommand{\ee}[1]{\label{#1} \end{equation}}  
\newcommand{\bee}{\begin{eqnarray}}             
\newcommand{\eee}{\end{eqnarray}}               

\begin{document}
\title{Polynomial law for controlling the generation of n-scroll chaotic attractors in an optoelectronic delayed oscillator}
\author{Bicky A. M\'{a}rquez}
\email[Email:~]{bmarquez@ivic.gob.ve}
\affiliation{Centro de F\'{i}sica, Instituto Venezolano de Investigaciones
Cient\'{i}ficas, km. 11 Carretera Panamericana, Caracas 1020-A, Venezuela}
\author{Jos\'{e} J. Su\'{a}rez-Vargas}
\email[Email:~]{jjsuarez@ivic.gob.ve}
\affiliation{Centro de F\'{i}sica, Instituto Venezolano de Investigaciones
Cient\'{i}ficas, km. 11 Carretera Panamericana, Caracas 1020-A, Venezuela}
\author{Javier A. Ram\'{i}rez}
\affiliation{Centro de F\'{i}sica, Instituto Venezolano de Investigaciones
Cient\'{i}ficas, km. 11 Carretera Panamericana, Caracas 1020-A, Venezuela}
\date{\today}
\begin{abstract}
Controlled transitions between a hierarchy of n-scroll attractors are
investigated in a nonlinear optoelectronic oscillator. Using the system's
feedback strength as control parameter, it is shown experimentally the
transition from Van der Pol-like attractors to 6-scroll, but in generall this
scheme can produce an arbitrary number of scrolls. The complexity of every state
is characterized by Lyapunov exponents and autocorrelation coefficients.
\end{abstract}
\maketitle


\textbf{Delayed nonlinear differential equations have been implemented
in electro-optical systems to generate highly complex optical chaos. These
implementations in turn allowed the construction of optical communication
systems with different degrees of encryption and security. However, many of
those were demonstrated to be breakable by different methods of signal
processing and computational analysis. Therefore, before embarking a new chaotic
system in such enterprize, it would be of primary interest to characterize the
dynamics using methods of phase space analysis that show hidden patterns
existing in complex structures. In this report we show experiments and analysis
of a time-delayed electro-optical feedback oscillator that generates an
arbitrary n-scroll attractor family in a controlled fashion. This method grants
a new degree of freedom when designing highly secure communication schemes, in
particular it allows fine-tuning and matching the system phase-space structure
with required complexity measures that maximize the security properties of a
communication scheme suitably designed.}

\section{INTRODUCTION}
Since the discovery of chaotic dynamics there has been a strong interest in
finding coherent structures in systems generating very complex patterns. From
the classic single\cite{Rossler} and 2-scroll attractors
\cite{Lorenz, Matsumoto, Brown}, many reports have shown systems capable of
displaying hyper-chaotic n-scroll dynamics\cite{Yu, Elhadj}. A number of those depend
on the existence of nonlinear, non-invertible functions in their mathematical
models. We highlight the work of Tang \textit{et al.}\cite{Tang} that replaces
the typical nonlinear piecewise-defined function by a sine function in the
Chua's circuit to generate multi-scroll attractors, and Zhong \textit{et al.}
\cite{Zhong} who also modified the nonlinear piecewise function in Chua's
circuit to generate 10-scroll chaotic attractors. In another example Yal\c{c}in
\textit{et al}. \cite{Yalcin} use a sine function as the nonlinear component to
generate n-scroll chaotic attractors on Josephson junctions systems.

Differential equations with time delay comprise another mechanism to generate hyper-chaos\cite{Sprott97}. In 1977 Mackey and Glass\cite{Mackey} proposed a
first-order delayed nonlinear differential equation, which describes a
physiological control system, that generates multi-scroll dynamics. Since then
a significant amount of hyperchaotic systems have been defined using these
equations, e.g. Wang \textit{et al.}\cite{Wang},  Sprott\cite{Sprott1} and
Yal\c{c}in \textit{et al} \cite{Yalcin2}.

In this Letter, we present a general n-scroll chaotic system generated by a
nonlinear delayed electro-optical feedback circuit. To build this circuit we
were inspired by a modified version of the Ikeda system \cite{Chembo, Peil}, and, more
recently, by an electro-optic circuit able to generate a variety of dynamics,
from regular to high-dimensional chaos\cite{Murphy}.

In a previous work \cite{Suarez} we showed that this system has the capability
to control the complexity of the dynamics by physical parameters other than the
delay. In this work, we show that the law for generating n-scroll is directly
related, and hence can be suitably controlled, by the number of peaks and
troughs of the nonlinear, non-invertible function part of the system's transfer
characteristic. Both experiments and theory supporting this discovery are
detailed.

\section {DELAYED NONLINEAR DIFFERENTIAL EQUATIONS}
\begin{figure}[h]
\includegraphics*[scale=0.28]{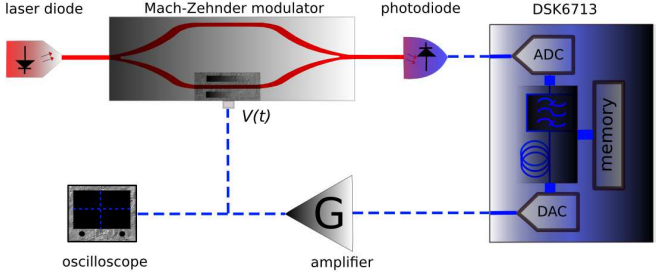}
\caption{\label{schem1_fig1} Schematic diagram of the delayed feedback optical
circuit that generates n-scroll hyper-chaos. It comprises a laser diode, a
Mach-Zehnder modulator, a photodiode, a DSK6713 DSP board and an amplifier. The
time-delay $T$ is $10.4 \mu s$, and the second-order bandpass digital filter is
constructed with a low-pass filter with cutoff frequency $f_{l}=1Hz$ and a
high-pass filter with cutoff frequency $f_{h}=10kHz$.}
\end{figure}

Fig. \ref{schem1_fig1} we show the delayed nonlinear electro-optical feedback
system \cite{Suarez}. It was implemented using a laser diode, a Mach-Zehnder
electro-optic modulator (MZM), a photodiode, an electronic amplifier, optical
fiber and a digital signal processing board (DSP) TMDSDSK6713 from Texas
Instruments. The circuit is modeled by a second-order delayed nonlinear
differential equation, or by a system of first-order delayed nonlinear
differential equations \cite{Peil}. The nonlinear transformation is implemented
by the MZM, whose transmission function is a cosine-squared static
characteristic. The modulator transforms the infrared beam, arriving from the
laser through a single-mode optical fiber. The photodiode converts the output of
the MZM into an electrical signal, which is time-delayed and digitally filtered
with the DSP.  The analog output voltage of the DSP is then amplified and
connected back to the MZM radio-frequency input, $V(t)$, to form the closed
loop.

The system of dimensionless first-order delayed nonlinear differential equations
is \cite{Chembo}:
\begin{equation}
\label{eq:1stODSyst}
\begin{array}{l}
\dot{x}=-x -\kappa \nu -\beta \cos^{2}(x_{\tau} + \phi_{0}),  \\
\dot{\nu}=x; \end{array}
\end{equation}
where,
\begin{equation}
\label{eq:1stODSyst_param}
\begin{array}{ll}
 x(t)=\pi V(t)/ 2V_{\pi}, & x_{\tau}=x(t-\tau), \\
\tau=2\pi Tf_{h}, & \beta=\pi \eta G S P_{in}/ 2V_{\pi},\\
\kappa = f_{l}/f_{h}.& \\
 \end{array}
\end{equation}
The parameters are the following: dimensionless time-delay $\tau$, feedback loop
gain $\beta$, the angle describing the bias point of the modulator $\phi_{0}$,
sensitivity of the photodiode $S$, half-wave voltage of the MZM $V_{\pi}$, $G$
is the gain of the amplifier circuit, the line attenuation factor $\eta$ and
$f_{h}, f_{l}$ are cutoff frequencies of the digital bandpass filter. The
experimental values are $f_{l}=1Hz$ and $f_{h}=10kHz$, $G=5.5$, $V_{\pi}=3.0 V$, $\phi_{0}=\pi/4$, $\tau=0.65$ and the laser power is varied from $104.4 \mu W$
to $1.76mW$.

\section{TIME-DELAYED LI\'{E}NARD AND VAN DER POL MODELS}
To better understand the topologies of the n-scroll in the phase space we make
analogies with two important mathematical systems. We begin our modeling by
supposing that initially the system has no time-delay, then the planar system of
Eq. (\ref{eq:1stODSyst}), with $\phi_{0}=\pi/4$ and $\tau=0$, becomes
\begin{equation}
\label{eq:delay0}
\ddot{x} + [1-\beta \cos(2x)]\dot{x} + \kappa x=0,
\end{equation}
This equation shows that the electro-optical feedback circuit without time-delay
is equivalent to a general Li\'enard system\cite{Lienard} described by the
equation,
\begin{equation}
\label{eq:lienard}
\ddot{x} + f(x)\dot{x} + g(x)=0,
\end{equation}
where $f(x)= 1-\beta \cos(2x)$ is the even function that models the damping and
$g(x)= \kappa x$ is the odd function describing the restoring force.

To find the Li\'enard plane associated to Eq. (\ref{eq:1stODSyst}), it is
necessary to derive the odd function $F(x)$ from the damping potential
$f(x)$. The Li\'enard plane is thus defined as \cite{Lynch}:
\begin{equation}
\label{eq:lienardplane}
\begin{array}{l}
\dot{x}=y - F(x),  \\
\dot{y}=-g(x); \end{array}
\end{equation}
where $F(x)= \int^{x}_{0} f(x')dx'$.
Therefore, the Li\'enard plane corresponding to Eq. (\ref{eq:delay0})  is,
\begin{equation}
\label{eq:lienardelay0}
\begin{array}{l}
\dot{x}=y - x + \beta \sin(2x)/2,  \\
\dot{y}=-\kappa x; \end{array}
\end{equation}
where the odd function $F(x)= x - \beta \sin(2x)/2$.
Expanding $F(x)$ in a sine Taylor series, and keeping only the first two terms,
we get a Van der Pol type oscillator \cite{Lynch}:
\begin{equation}
\label{eq:VdP}
\begin{array}{l}
\dot{x}=y - (1-\beta)x + 2\beta x^{3}/3,  \\
\dot{y}=-\kappa x; \end{array}
\end{equation}
whose corresponding potential function is given by
$f(x)=1-\beta + 2\beta x^{2}$.

\begin{figure}[h]
\includegraphics*[scale=0.45]{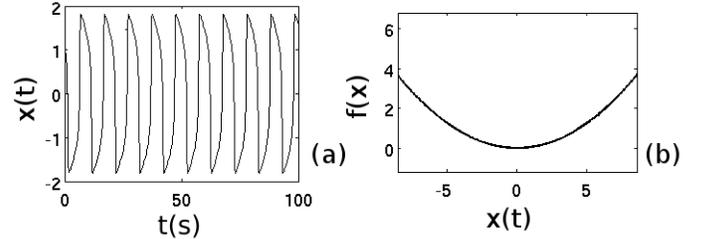}
\caption{\label{fig2} (a) Numerical solution, $x(t)$, of the Van der Pol type
oscillator of Eq. (\ref{eq:VdP}), and (b) the associated potential function
$f(x)$.}
\end{figure}

The numerical solution $x(t)$ of Eq. (\ref{eq:VdP}) and the potential function
$f(x)$ are shown in Fig. (\ref{fig2}). The potential has a parabolic shape,
which means that the dynamics is bounded on a single-well. Mono-stable
potentials in systems with two degrees of freedom indicate the existence of a
limit cycle, corresponding to periodic orbits.

\begin{figure}[h]
\includegraphics*[scale=0.45]{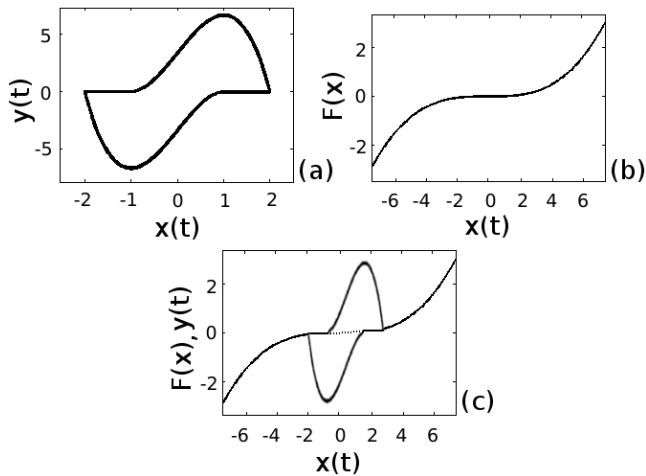}
\caption{\label{fig3} (a) Phase portraits of the Van der Pol oscillator in the
Li\'enard plane. (b) The odd  function $F(x)$, and (c) overlapping the preceding
figures.}
\end{figure}
The Li\'enard plane of the Van der Pol oscillator, Fig. (\ref{fig3}), shows
direction and flow velocity, and how it is affected by the shape of the
potentials. The range in which the two curves are superimposed,
Fig. (\ref{fig3}-c), constitutes the epochs where the time series changes
slowly, whereas the upper and lower curved protuberances of the phase plane,
form fast jumps from a region to the other.
The advantage of writing general systems in variables of Li\'enard allows us to
directly observe the effect of functions associated to potentials in
experimental physical systems like the optoelectronic delayed oscillator.
It is well-known that a planar system, in the absence of the time-delay, cannot
display chaos. By adding the time-delay, $\tau \neq 0$, there is an increase on
the dimensionality of the planar system. We now write the time-delayed version
of Eq. (\ref{eq:delay0}) that truly represents our original system model
(\ref{eq:1stODSyst}):
\begin{equation}
\label{eq:delayed}
\ddot{x} + \dot{x} -\beta \cos(2x_{\tau})\dot{x}_{\tau} + \kappa x=0,
\end{equation}
with Li\'enard plane:
\begin{equation}
\label{eq:lienardelayed}
\begin{array}{l}
\dot{x}=y - x + \beta \sin(2x_{\tau})/2,  \\
\dot{y}=-\kappa x. \end{array}
\end{equation}

\section{N-SCROLL CHAOTIC ATTRACTORS FAMILY}
\begin{figure}[h]
\includegraphics*[scale=0.45]{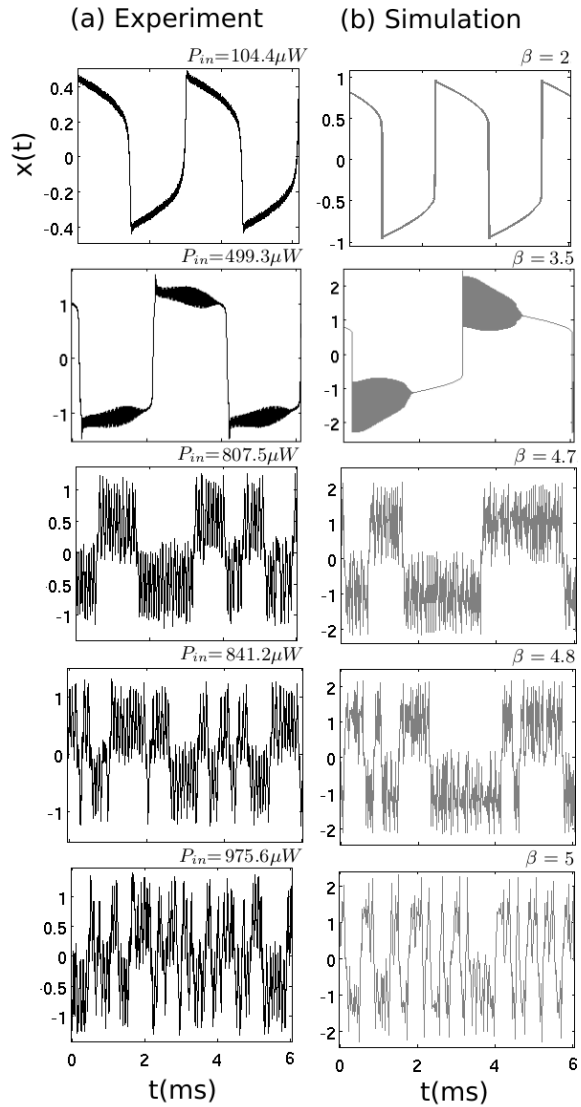}
\caption{\label{fig4} Time series of the delayed electro-optical feedback
system: (a) experiments and (b) simulations of Eq. (\ref{eq:delayed}).
The laser power was changed in the values $P_{in}=\{ 104.4\mu W, \; 499.3\mu W,
\; 807.5\mu W$, $841.2\mu W$ and $975.6\mu W \}$. The simulation parameter was
varied as $\beta=\{ 2, \; 3.5, \; 4.7, \; 4.8$ and $5 \}$.}
\end{figure}
We now use the model Eq. (\ref{eq:delayed}) to perform the simulations and guide
the experiments. In this implementation we fixed the time-delay at
$T=10.4 \mu s$, and show the route to chaos, in Fig. (\ref{fig4}), as we vary
the laser power $P_{in}$, in the experiments, and $\beta$ in the numerical
simulations. In this way we found a number of interesting bifurcations: the
first bifurcation is a periodic time series occurring at $P_{in}= 104.4\mu W$,
which has exactly the same shape as the Van der Pol oscillator shown in
Fig. (\ref{fig2}); in the next bifurcation, at $P_{in}= 499.3\mu W$, emerges the
so-called chaotic breathers \cite{Chembo}, they constitute the onset of chaotic
oscillations \emph{breathing} periodically; and finally we found a series of
bifurcations showing chaos and n-scroll hyper-chaos.

\begin{figure}[h]
\includegraphics*[scale=0.45]{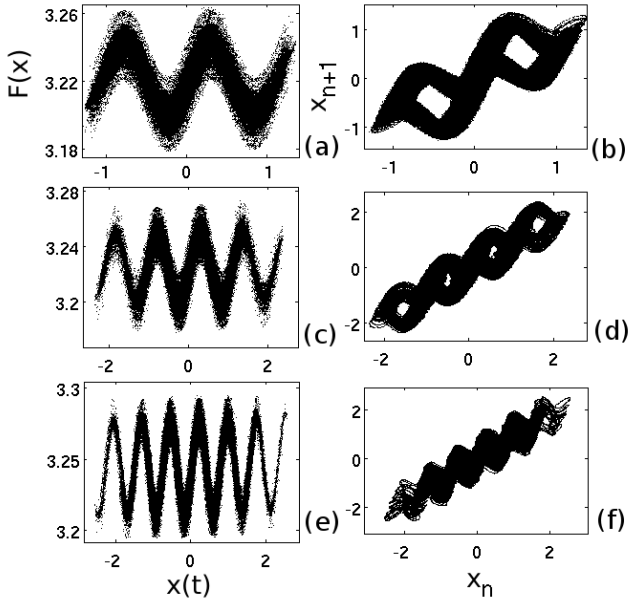}
\caption{\label{fig5} (a,c,e) Transmission functions of the MZM,
and (b,d,f) first return maps from the time series. In these experiments the
values of laser power were $P_{in}=\{807.5\mu W$, $975.6\mu W$ and $1.76mW \}$.}
\end{figure}
Increasing the laser power to $P_{in}=807.5\mu W$, Fig. (\ref{fig5}-a),
increases the number of peaks and troughs of the MZM transmission function
$\cos^{2}$. This results in a polynomial whose degree describes the potential
function of Eq. (\ref{eq:delayed}),
$f(x)=1-\beta + 2\beta x^{2} - 2\beta x^{4}/3$; therefore, the odd variable of
Li\'enard is a polynomial of degree five in Eq. (\ref{eq:lienardelayed}),
$F(x)=1-\beta x + 2\beta x^{3}/3 - 2\beta x^{5}/15$. Because the potential
function $f(x)$ is quartic, we expect to find a double-well potential that is
irregularly visited by the chaotic orbits, Fig. (\ref{fig5}-b), forming a
2-scroll chaotic attractor. When the laser power is $P_{in}=975.6\mu W$ and
$P_{in}=1.76mW$, we obtain polynomials of degree nine and thirteen respectively,
Fig. (\ref{fig5}-c,e). In Fig. (\ref{fig5}-d,f) we show experiments
corresponding to potential functions formed by polynomials of degree eight and
twelve, i.e. four-fold and six-fold potential wells. This results in 4-scroll
and 6-scroll chaotic attractors.

\begin{figure}[h]
\includegraphics*[scale=0.45]{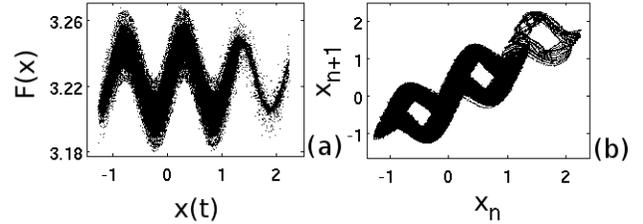}
\caption{\label{fig6} (a) Transmission function of the MZM and (b) the first
return map from the time-series. Laser power $P_{in}=841.2\mu W$.}
\end{figure}

We notice in Fig. (\ref{fig6}) that the frequency of visits to potential wells
depends on whether they are well-defined, i.e., because the polynomial is not
well-formed at one of its ends, results in trajectories not visiting that region
too often. Then the corresponding scroll will be less filled than in the other
cases. We note also that the number of scrolls that can be obtained from this
system will be greater the larger the degree of the polynomials are realized
experimentally on the MZM. This poses a technological challenge for the
implementation of arbitrarily large number of scrolls, and at the same time
motivates the design of new systems with nonlinear characteristics that can
provide this added benefit. In our particular setup using a JDSU-21014994
MZM we can only get up to a polynomial of degree thirteen, i.e., sixfold-well
potential, realizing a six-scroll attractor.
\section{COMPLEXITY INDICES: LYAPUNOV EXPONENT AND AUTOCORRELATION COEFFICIENT}
\begin{figure}[h]
\includegraphics*[scale=0.6]{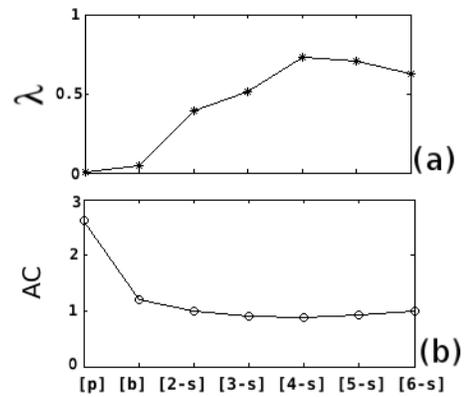}
\caption{\label{fig7} Complexity indices: (a) Lyapunov exponents, $\lambda$, and (b) autocorrelation coefficients, AC, for the periodic [p], breathers [b], 2-scroll [2-s], 3-scroll [3-s], 4-scroll [4-s], 5-scroll [5-s] and 6-scroll [6-s] chaotic attractors. }
\end{figure}

Besides being able to discern the structure of the phase space, it is always
useful to classify the attractors in terms of indices that summarize the
predictability of the time series. In Fig. (\ref{fig7}) we show the maximum
Lyapunov exponent and autocorrelation coefficient that were calculated using the
software package \texttt{TISEAN} \cite{Hegger}, for each of the n-scroll
attractor generated by the electro-optical system. Both Lyapunov exponent,
Fig. (\ref{fig7}-a), and correlation coefficient, Fig. (\ref{fig7}-b), behave
monotonously, increasing the first and decreasing the latter, until the
4-scroll attractor. This means, the complexity in the system is growing
proportionally with laser power. However, for 5-scroll the system unexpectedly
behaves more predictable, breaking the monotony in the dependence of complexity
with laser power. This behavior was observed also, for instance, in Murphy
\textit{et al}. \cite{Murphy}, where the break of monotonous behavior was
evident when the feedback intensity of their electro-optical system was
increased, from $P_{in} = 841.2\mu W$.
\section{CONCLUSION}
For a well-known optoelectronic oscillator we presented a polynomial law to
predict and control the generation of n-scroll chaotic attractors, based only
on the system's nonlinear static characteristic. Using the Li\'enard equations
and phase space analysis we were able to model this electro-optical delayed
feedback system, and discovered the polynomial law that generates n-scroll
chaotic attractors: the polynomial functions, $F(x)$, of degree $(N+1)$
performed by the MZM, are associated to Li\'enard potentials, $f(x)$, described
by polynomials of degree $(N)$, which correspond to (N/2)-scroll chaotic
attractors. Importantly we associated complexity indices for each attractor
structure and determined that there is a break in monotonous behavior of
complexity indices Vs. laser power, as has been corroborated in other works.
\acknowledgments
This work was supported under IVIC project $N^{\circ}$ 448: ''Nonlinear dynamics in
biological and technological systems''.

%
\end{document}